\documentclass[
 reprint,
 amsmath,amssymb,
 aps,
]{revtex4-2}

\usepackage{amsmath}
\usepackage{bm}
\usepackage{graphicx}
\usepackage{amssymb}
\usepackage{times}
\usepackage{color}
\usepackage{float}
\usepackage{multirow}
\usepackage{subcaption}
\usepackage{braket}
\usepackage{hyperref}
\usepackage{chemformula}
\usepackage{makecell}
\usepackage{xfrac} 
\usepackage{threeparttable} 
\usepackage{placeins}

\begin{document}
\raggedbottom
\title{Sorption Molecular Dynamics of Point Defects in 2D Materials with a Machine Learning Potential}

\author{Nils-Ruben Skogemyr}
\thanks{These authors contributed equally.}
\author{Albin Sjöstedt}
\thanks{These authors contributed equally.}
\author{Isak Wernroth}
\thanks{These authors contributed equally.}
\author{Isak Häggqvist}
\thanks{These authors contributed equally.}
\author{Isac Ljungwald}
\thanks{These authors contributed equally.}
\author{Hans-Ove Åkesson}
\thanks{These authors contributed equally.}

\affiliation{Department of Physics, Chemistry and Biology, Link\"oping University, Link\"oping, Sweden}

\author{Henricus R.A. ten Eikelder}
\affiliation{Department of Physics, Chemistry and Biology, Link\"oping University, Link\"oping, Sweden}

\author{Rickard Armiento}
\affiliation{Department of Physics, Chemistry and Biology, Link\"oping University, Link\"oping, Sweden}

\author{Joel Davidsson}
\email{joel.davidsson@liu.se}
\affiliation{Department of Physics, Chemistry and Biology, Link\"oping University, Link\"oping, Sweden}

\begin{abstract}

Defect engineering is a crucial tool for tuning and tailoring the properties of two-dimensional (2D) materials.
A key challenge is identifying which defects are kinetically stable at room temperature and sufficiently promising to warrant further investigation.
However, identifying stable defects with suitable properties using first-principles calculations, such as density functional theory (DFT) is computationally demanding, effectively limiting the scope of high-throughput studies. 
In this work, we investigate whether the machine learning interatomic potential MACE can reproduce sorption trends (adsorption vs. absorption) previously obtained from high-throughput DFT calculations and extend the analysis to elevated temperatures.
A total of 17,624 defect structures from the ADAQ database are simulated using MACE-accelerated molecular dynamics (MD) at both 0~K and 300~K.
The results show that MACE successfully reproduces the overall sorption trends observed in DFT calculations at 0 K at a fraction of the computational cost.
Simulations performed at 300 K further indicate that the qualitative sorption trends remain largely the same when including finite-temperature effects.
These findings demonstrate that MACE-driven MD can be used as an effective high-throughput prescreening tool for defect-related studies in 2D materials, specially related to sorption. 
While conventional first-principles methods, such as DFT, remain necessary for accurate and quantitative validation, the use of MLIPs enables MD simulations of finite-temperature defect behavior at a fraction of the computational cost of DFT-based MD.

\end{abstract}

\maketitle

\section{Introduction}

Identifying promising defects in two-dimensional (2D) materials is central to the development of sustainable materials~\cite{Zhuang2025, Xiang2025, Krasheninnikov2026}. 
These materials are particularly attractive because of the unique and highly tunable properties stemming from the reduced dimensionality~\cite{Krasheninnikov2026}. 
Defect engineering provides a versatile route to tailoring their electronic, optical, chemical, and structural properties~\cite{Krasheninnikov2026,Efat2026}.
One such approach involves introducing dopant atoms through sorption, the process by which atoms or molecules are taken up by a material through either absorption or adsorption. 
Here, we investigate whether single dopant atoms occupy interstitial sites within the 2D material (absorption) or bind to its surface (adsorption).

Identifying stable defects is challenging due to numerous possible defect configurations, as small structural changes can significantly affect their stability and properties~\cite{Igumbor2022, Gastaldi2026, Deng2025}. 
Simulations enable high-throughput computational screening of potential defects in 2D materials to evaluate stability and identify promising candidates efficiently, reducing the time and cost of experimental screening~\cite{Sluydts2016, xiong2023}.
A commonly used approach for determining defect properties is density functional theory (DFT)~\cite{KohnSham1965, hohenberg1964inhomogeneous}, which enables accurate predictions for many material properties. 
Although static calculations of relatively small systems are computationally manageable, the repeated energy and force evaluations needed for molecular dynamics (MD) can become computationally intensive.
As a result, MD~\cite{FrenkelSmit2023} based on DFT forces are typically limited to systems containing a few hundred atoms and short time scales~\cite{Baghishov2026,Zhang2026ML,Arabha2021}. 
Using classical interatomic potentials is a computationally efficient approach capable of simulating much larger systems over longer time scales than DFT.
However, when using classical potentials (e.g., Lennard-Jones and Stillinger-Weber), the accuracy falls short of that achieved with DFT~\cite{Wang2024}. 

Machine learning interatomic potentials (MLIPs) offer a solution to this trade-off, bridging the gap between accuracy and efficiency~\cite{Dicks2026}. 
These models are usually trained on large datasets containing ab initio calculations, such as the Materials Project (MP) platform ~\cite{Horton2025}, enabling near-DFT accuracy at a fraction of the computational cost~\cite{jacobs2025}.
One of these models, and the one used in this study, is the Machine Learning Force Fields (MACE) model. 
The MACE architecture is based on higher-order equivariant message passing, enabling an accurate description of many-body atomic interactions while preserving physical symmetries such as rotations and reflections~\cite{https://doi.org/10.48550/arxiv.2206.07697}. 
The explicit inclusion of higher-order correlations between neighboring atoms, allows the model to achieve high accuracy with only two message passing steps, resulting in improved computational efficiency. 
Like other similar models, MACE has been trained on a significant number of DFT calculations and is known to perform well on monolayer and 2D materials~\cite{Sauer2025, batatia2025foundationmodelatomisticmaterials}.

Recent studies have demonstrated the applicability of MACE and similar MLIPs across a broad range of materials science problems. 
The MACE MLIP enables efficient and accurate high-throughput harmonic phonon calculations for evaluating vibrational properties, facilitating large-scale screening of material stability that would be computationally demanding with conventional DFT~\cite{Lee2025}.
This potential has also been used to calculate formation energies for point defects in 2D materials~\cite{Berger25}. 
Other MLIPs have been applied to accelerate large-scale defect exploration and finite-temperature simulations of material properties, including melting temperature calculations and studies of 2D materials~\cite{Liu2025,Li2026,Hong2026}. 
They have also been used for high-throughput screening of dopants in transition metal dichalcogenides~\cite{Faiyad2026, Bhawsar2024}.
However, the applicability of these models to more complex defect processes and mechanisms, such as adsorption and absorption, remains largely unexplored. 

Previous high-throughput DFT calculations produced a database showing that the preferred dopant location, whether it adsorbs on the surface or diffuses into the layer, depends on dopant and material properties~\cite{abad}.
A small subset of the defects from this database have been studied with MLIPs at both 0 K and finite-temperature~\cite{Kretschmer2026}.
However, these prior studies have not investigated how these trends evolve at finite temperatures across a large number of defect structures. It is also not established if the trends are reproduced by foundation-model MLIPs without fine-tuning, which would allow obtaining them at a much reduced computational expense compared to DFT.

In this work, we show that the MACE potential indeed reproduces the general trends at 0 K observed in DFT calculations~\cite{abad}.
We also investigate finite-temperature effects of thousands of defect structures using MACE-driven MD, which would be computationally unfeasible with DFT-based MD.
Hence, we show, at a significantly reduced computational cost, that MACE reproduces the overall sorption trends at 0 K, as well as largely conserves the qualitative trends at 300 K.

\section{Method}

\begin{figure*}[tp]
    \centering
    \includegraphics[width=\textwidth]{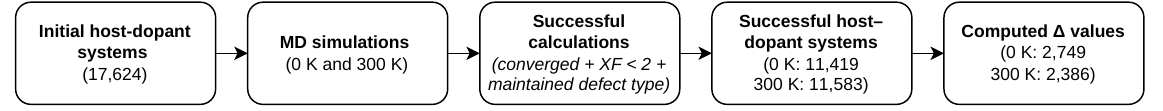}
    \caption{Workflow used to construct the final dataset. Only host--dopant systems satisfying the energy and expansion factor criteria for both adsorption and absorption configurations were included in the final $\Delta$ analysis.}
    \label{fig:filtering_pipeline}
\end{figure*}

The MD calculations are carried out using
the Atomic Simulation Environment (ASE) ~\cite{HjorthLarsen2017} and ASAP3 Python libraries \cite{asap3}. Interatomic interactions are modeled using the MACE potential, with the \textit{mace-mpa-0-medium.model} ~\cite{batatia2025foundationmodelatomisticmaterials}. This model is trained on the Materials Project Trajectory (MPTrj) \cite{mptraj} and subsampled Alexandria (sAlex) \cite{BarrosLuque2026} datasets, where both are based on DFT calculations using the PBE exchange-correlation functional with a Hubbard \(U\) correction (PBE+U)~\cite{Perdew1996,Dudarev1998}. To manage the large-scale calculations, the high-throughput toolkit (\textit{httk})~\cite{httk} is used to automate the task generation and execution. The 17,624 structures used in the calculations are taken from the imp2D database~\cite{abad}, which have been incorporated into the ADAQ database ~\cite{DAVIDSSON2021108091,adaqdatabase}.
Each of these simulated systems consists of a single dopant in a 2D host material. The dopant is initially placed at either an adsorption site on the surface or an interstitial site within the  material. The same initial structures as in Davidsson et al.~\cite{abad} are employed to allow for direct comparison between the MD and DFT methods.

Each system is simulated in the NVT ensemble at 0 K and 300 K, using the Langevin integrator. The friction parameter used in the integrator is \(\gamma = 0.05\) in both cases.
The convergence criterion employed for the 0 K case verifies that the energy difference \(\Delta E\) satisfies \(\Delta E \leq 0.01\ \mathrm{eV}\), where \(\Delta E = \lvert E_{max} - E_{min} \rvert \) is calculated over the last 100 MD steps. 
In the case of the 300 K simulations, due to high energy fluctuations caused by the stochastic temperature insertion of the Langevin integrator, a new convergence criterion based on floating mean error is implemented.
\begin{equation}
    \Biggl| \frac{\bar E_{tot}^{[n-500, n]} - \bar E_{tot}^{[n-1000, n-500]}}{\bar E_{tot}^{[n-500, n]}} \Biggl| < \frac{10^{-4}}{\sqrt{N}}.
\end{equation}
Where \(n\) denotes an arbitrary step within the simulation interval, \(E_{tot}\) is the total energy of the system, and \(N\) denotes the total number of atoms in the simulated cell.
A maximum number of steps is set to terminate simulations of materials that exhibit quasi-stable behavior and elude the convergence- and divergence-criterion. For both 0 K and 300 K, the limit is set to a maximum of 10,000 steps of 0.5 fs.

The same classification parameters for a defect of type $X$ in the host material $H$ as in Ref.~\onlinecite{abad} were used. The depth parameter $D$ is used to classify the defect site as either an adsorption or an interstitial site. 
\begin{equation}
    D[H,X] = \frac{2z[X] - (z_{min}[H]+z_{max}[H])}{z_{max}[H]-z_{min}[H]} ,
\end{equation}
where $z[X]$ denotes the z-coordinate of the defect, while $z_{min}$ and $z_{max}$ describe the minimum and maximum z-coordinate for any atoms in the crystal. Expansion factor XF is defined as

\begin{equation}
    XF [H,X] = \frac{d_{rel}[H,X]}{d_{unrel}[H,X]},
\end{equation}
where $d_{unrel}$ is the thickness of the 2D material before relaxation and $d_{rel}$ the thickness of the system after relaxation. The metric used to decide if an absorbed or adsorbed state is preferred is $\Delta$, defined as
\begin{equation}
    \Delta[H,X] =E_{int}^{f,min}[H,X] - E_{ads}^{f,min}[H,X], 
\end{equation}
where $E_{int}^{f,min}$ denotes the lowest formation energy found for an interstitial defect, and $E_{ads}^{f,min}$ is the same but for an adsorbed defect. Formation energy at 0 K is based only on the last converged time step, while at 300 K it uses the mean of the last 500 time steps to minimize uncertainty from thermal fluctuations.

A computation is considered successful if it converges in total energy, satisfies the expansion factor criterion ($XF<2$), and maintains the same defect type. Only successful results are used in the analysis. The workflow used to construct the final dataset is detailed in Fig.~\ref{fig:filtering_pipeline}. Valid $\Delta$ values are calculated only for host--dopant systems for which both the adsorption and absorption configurations satisfy all convergence and classification criteria.

\section{Results}

In this section, we first compare the sorption trends of MACE with DFT at 0 K. 
The calculations require approximately 2,500 core-hours, which is significantly less (about 0.13\%) than the over 2 million core-hours needed for the original DFT study \cite{abad}.
Then we present the sorption trends observed at 300 K using MACE. 
Of the 17,624 investigated structures, 11,419 satisfy the convergence and classification criteria at 0 K, and 11,583 in the final 300 K dataset. The valid structures generate 2,749 $\Delta$ values at 0 K and 2,386 at 300 K.

\subsection{0 K - MACE vs. DFT}

\begin{figure*}
\includegraphics[width=0.9\textwidth]{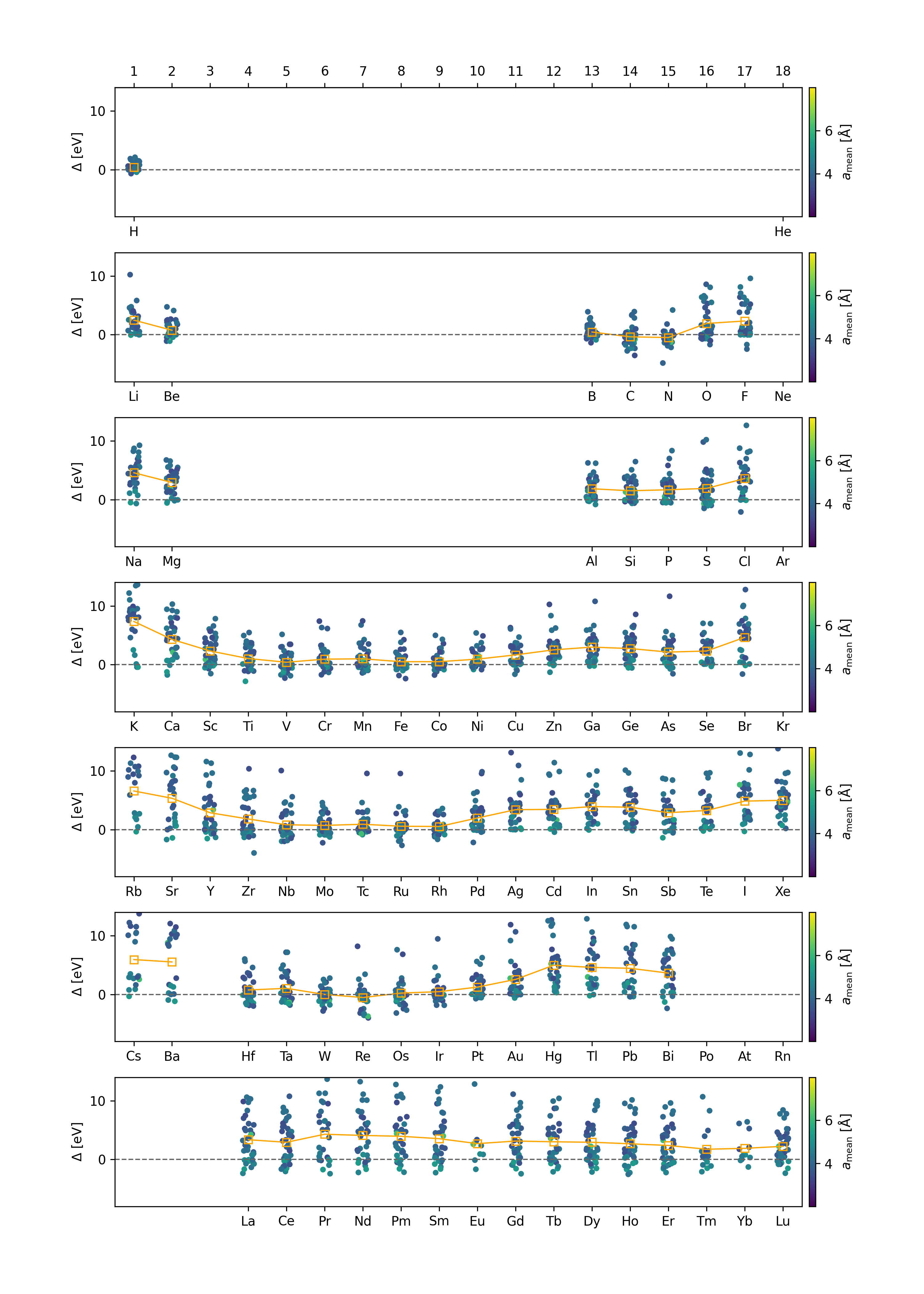}
    \caption{MACE-calculated $\Delta$ values at 0 K for all investigated defect--host combinations. Positive values indicate that adsorption is energetically preferred over absorption, while negative values indicate the opposite. The dopants are ordered according to the periodic table, in the same style as in Fig.~3 in Ref.~\onlinecite{abad}.}
    \label{fig:MACEDelta} 
\end{figure*}

The recreated $\Delta$-plot using MACE is shown in Fig.~\ref{fig:MACEDelta}. The plot shows the difference between the lowest interstitial and adsorption formation energies for each defect-host combination. Positive values indicate a preference for adsorption, while negative values indicate a preference for absorption.

Our results reproduce the overall periodic trends reported in Fig.~3  of ~\citet{abad}. Positive $\Delta$ values dominate throughout the dataset, indicating that adsorption is generally preferred over absorption. Large positive values are commonly observed for alkali and alkaline-earth metal dopants, while many transition-metal dopants exhibit values closer to $\Delta = 0$, indicating a weaker energetic preference between adsorption and absorption.

\begin{figure}[H]
\centering
\includegraphics[width=\columnwidth]{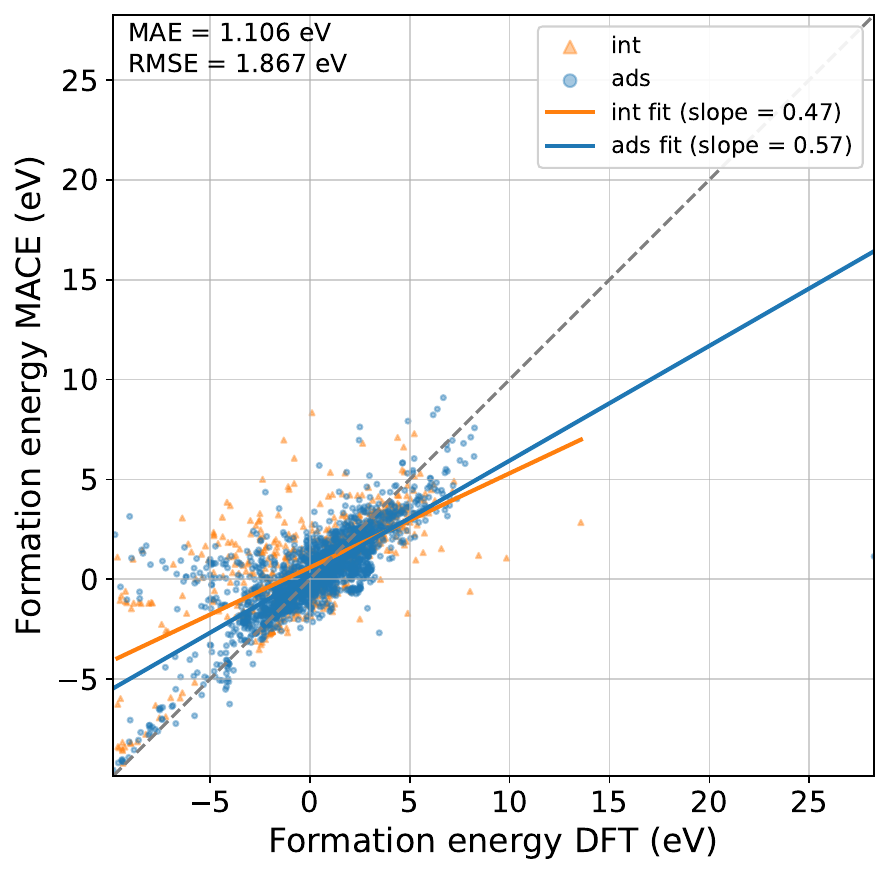}
\caption{Comparison of defect formation energies calculated using MACE and DFT. The black dashed line indicates perfect agreement between the two methods. 
The configurations for adsorption (ads), shown in blue, and absorption (int), shown in orange, are depicted separately. The mean absolute error (MAE) and root mean square error (RMSE) across all configurations are 1.106 eV and 1.867 eV, respectively. The blue and orange solid lines represent linear fits to the adsorption and absorption data, respectively. The fitted slopes of 0.57 and 0.47 indicate that MACE predicts a smaller variation in formation energies than DFT along the fitted trends for both configurations.
}
\label{fig:formationE}
\end{figure}

Although quantitative differences are observed between MACE and DFT formation energies, the same chemical trends are observed across the periodic table.
The relationship between the two methods can be examined in Fig.~\ref{fig:formationE}, where the formation energies obtained using MACE are compared directly with the DFT values. A clear positive correlation is observed, with an MAE of 1.106 eV and an RMSE of 1.867 eV. These values indicate that quantitative deviations remain between the two methods. However, the observed correlation and the reproduced adsorption--absorption trends demonstrate that MACE captures the relative energetic ordering of defect configurations.

\begin{figure*}  \includegraphics[width=0.9\textwidth]{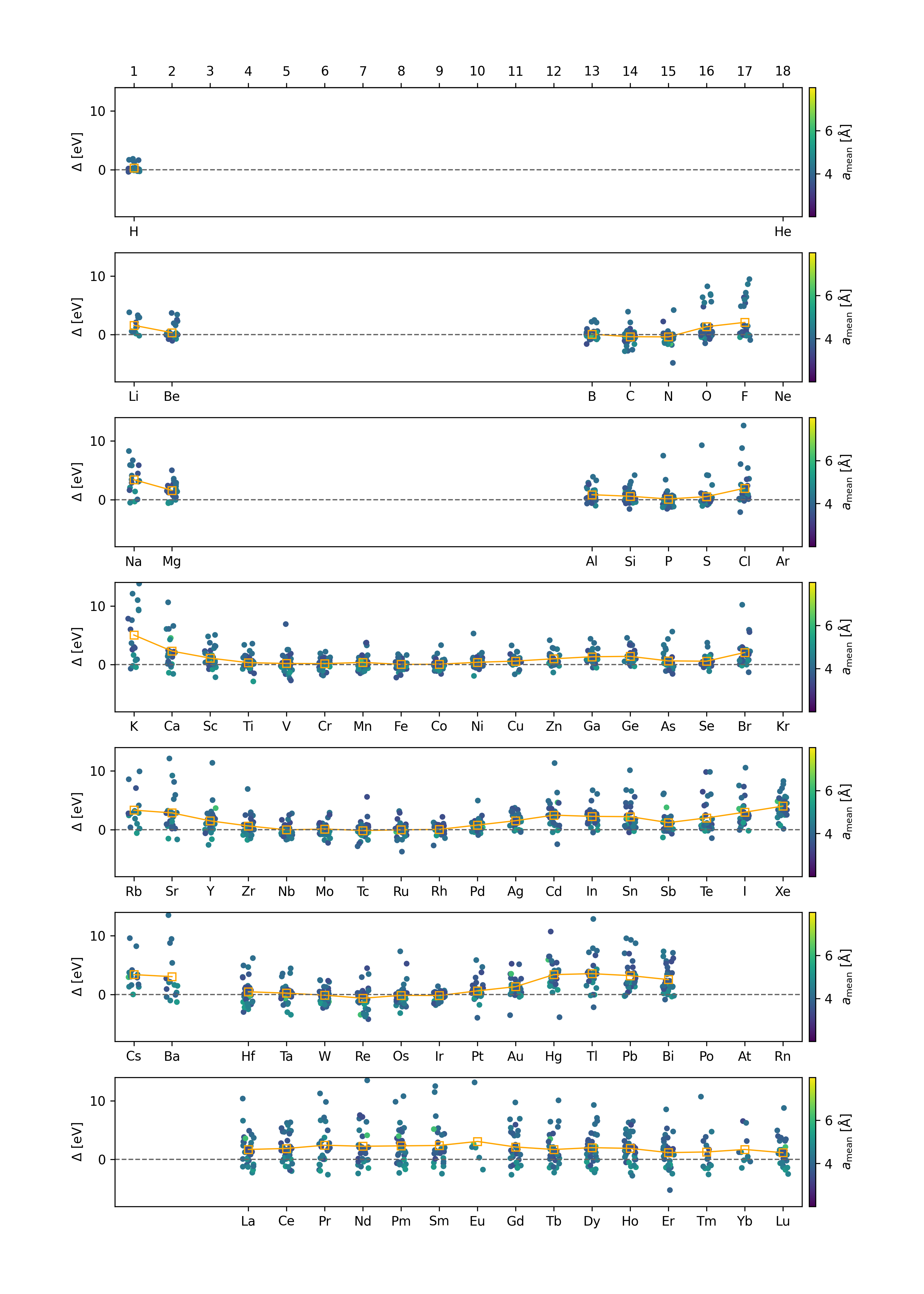}
    \caption{MACE-calculated $\Delta$ values at 300 K for all successfully converged defect--host combinations. Positive values indicate that adsorption is energetically preferred over absorption, while negative values indicate the opposite.}
    \label{fig:300KDelta} 
\end{figure*}

\subsection{300 K - MACE}
The relatively low computational costs of MACE enabled the extension of this study to finite-temperature simulations.
Figure~\ref{fig:300KDelta} shows the corresponding $\Delta$ values obtained from the 300 K simulations. 
The overall adsorption--absorption trends remain largely unchanged compared to 0 K, indicating that adsorption is still energetically preferred for the majority of the systems.
The main difference compared to the 0 K results is that the variation in $\Delta$ values is reduced for several groups of elements. Specifically, many dopants exhibit  $\Delta \approx 0$ at 300 K, suggesting that finite temperatures appears to reduce the energetic separation of adsorption and absorption configurations.

At 300 K, more systems satisfy the convergence and classification criteria. However, fewer adsorption--absorption pairs are obtained for the $\Delta$ calculations. The reduction in $\Delta$ values is not caused by poorer numerical convergence, but rather by thermally activated defect migration. 
In particular, some initially interstitial defects migrate out of the layer and become surface adsorbates during the simulation. As a result, the interstitial configuration transforms into an adsorption configuration, eliminating the adsorption--absorption pair, reducing the number of valid $\Delta$ values, and effectively moving the corresponding points towards $\Delta = 0$. The reduced number of valid adsorption--absorption pairs indicates that finite-temperature simulations can reveal additional instability mechanisms that are not captured in static 0 K calculations.

\section{Discussion}

We show that MACE can reproduce the sorption trends from DFT well, when comparing our Fig.~\ref{fig:MACEDelta} with Fig.~3 in Ref.~\onlinecite{abad}.
The strongest agreement between the MACE and DFT results is found in the qualitative trends rather than in exact formation energy values, as also indicated by the spread observed in Fig.~\ref{fig:formationE}. Although the exact formation energies may differ, MACE generally reproduces the DFT-predicted preference for adsorption or absorption. This difference in energy is expected, since the MACE model is a general-purpose MLIP trained on large DFT datasets rather than a model specifically fitted to the defect systems studied here.
The largely unchanged adsorption--absorption trends at 300 K suggest that the energetic differences governing the preferred defect configurations are sufficiently large for many systems to remain robust against moderate thermal fluctuations.

However, the reduced number of calculated $\Delta$ values at 300~K, illustrated in Fig.~\ref{fig:300KDelta}, show that finite-temperature simulations require careful classification procedures to ensure meaningful comparisons between adsorption and absorption configurations. 

The systems that fail the convergence or classification criteria may themselves provide useful information, as they are likely to correspond to shallow energy minima or metastable configurations that become unstable when thermal effects are introduced. In addition, some initially interstitial defects migrated toward the surface during the finite-temperature simulations, reducing the number of valid adsorption--absorption pairs available for the final $\Delta$ analysis.

These findings demonstrate that MACE can reliably reproduce qualitative adsorption--absorption trends while providing a practical compromise between computational efficiency and predictive capability, making the method well suited for large-scale screening of defect configurations prior to more computationally demanding DFT calculations.

\section{Conclusion}
The results show that MACE can reproduce the general DFT trends for adsorption versus absorption in doped 2D materials, while requiring only a fraction of the computational cost, making the method suitable as a high-throughput screening tool for identifying promising defect--host combinations before performing more expensive DFT calculations.

The overall sorption trends remain largely unchanged at 300 K despite thermal activation, indicating that the energetic preference for adsorption is robust for most investigated systems. At the same time, the finite-temperature simulations reveal thermally activated defect migration and other stability effects that cannot be captured by static 0 K calculations alone. The inclusion of finite-temperature effects enables the investigation of defect behavior under more realistic operating conditions and may reveal temperature-induced trends that are inaccessible in conventional high-throughput 0 K DFT studies, demonstrating the potential of machine-learning interatomic potentials for efficient large-scale finite-temperature studies of defects in 2D materials.

Overall, the presented approach demonstrates that machine-learning interatomic potentials can substantially expand the scope of computational defect studies while retaining the qualitative sorption trends obtained from DFT.

\section{Author contributions}
NRS, AS, IW, IH, IL, and HOÅ contributed equally to Investigation, Formal analysis, Data curation, Validation, Visualization, and Writing – original draft. HRAE: Methodology, Supervision, Project administration. RA and JD: Conceptualization, Methodology, Supervision, Project administration. All authors: Writing – review \& editing.

\section*{Acknowledgments}

The work presented in this paper stemmed from a project conducted in the undergraduate course \emph{Project Course in Applied Physics, CDIO} (TFYA99) at Linköping University. RA acknowledges support from the Swedish Government Strategic Research Area Swedish e-science Research Centre (SeRC).
J.D. acknowledges support from the Swedish Research Council (VR) Grant No. 2022-00276.
The computations were enabled by resources provided by the National Academic Infrastructure for Supercomputing in Sweden (NAISS), partially funded by the Swedish Research Council through grant agreement no. 2022-06725.

\newpage
\bibliography{references}

\end{document}